\documentclass[aps,twocolumn,prb,showpacs]{revtex4}
\usepackage{graphicx,amssymb,amsmath}
\usepackage{graphicx}
\usepackage{dcolumn}
\usepackage{bm}

\begin{document}

\title{Thermopower of Two-Dimensional Electrons at $\nu =$ 3/2 and 5/2}

\author{W.~E. Chickering$^1$, J.~P. Eisenstein$^1$, L.~N. Pfeiffer$^2$ and K.~W. West$^2$}

\affiliation{$^1$Condensed Matter Physics, California Institute of Technology, Pasadena, CA 91125
\\
$^2$Dept. of Electrical Engineering, Princeton University, Princeton, NJ 08544}

\date{\today}

\begin{abstract}
The longitudinal thermopower of ultra-high mobility two-dimensional electrons has been measured at both zero magnetic field and at high fields in the compressible metallic state at filling factor $\nu = 3/2$ and the incompressible fractional quantized Hall state at $\nu = 5/2$. At zero field our results demonstrate that the thermopower is dominated by electron diffusion for temperatures below about $T = 150$ mK. A diffusion dominated thermopower is also observed at $\nu = 3/2$ and allows us to extract an estimate of the composite fermion effective mass.  At $\nu = 5/2$ both the temperature and magnetic field dependence of the observed thermopower clearly signal the presence of the energy gap of this fractional quantized Hall state.  We find that the thermopower in the vicinity of $\nu = 5/2$ exceeds that recently predicted under the assumption that the entropy of the 2D system is dominated by non-abelian quasiparticle exchange statistics.

\end{abstract}

\maketitle

\section{\label{intro}Introduction}
The study of two-dimensional electron systems (2DES) in semiconductor heterostructures has yielded an astonishing number and variety of important physics results, both in theory and experiment.  When a magnetic field is applied perpendicular to a clean 2DES, the kinetic energy spectrum of the system is collapsed into a discrete set of highly degenerate Landau levels. At low temperatures and high magnetic fields, this quenching of the kinetic energy renders electron-electron interactions the dominant influence on the behavior of the system.  These interactions are directly responsible for the many remarkable incompressible fractional quantized Hall states and compressible composite fermion metallic phases which have been observed in high mobility 2DESs.

The great majority of the experimental studies of 2DESs in the high magnetic field regime have focussed on the electrical transport coefficients of the system, while a smaller number have addressed the elementary excitations of the 2DES using optical probes \cite{perspectives}.  In contrast, the thermal properties of strongly interacting 2DESs have received less scrutiny \cite{Gallagher,Fletcher}.

In this paper we examine the thermopower of 2D electron systems in ultra-pure GaAs/AlGaAs heterostructures at both zero and high magnetic field, $B$. Thermopower offers a perspective on 2DESs which is complementary to that provided by ordinary electrical transport.  For example, at $B=0$ the longitudinal thermopower, or Seebeck coefficient $S$, of the 2DES is sensitive to the energy derivative $\partial \tau /\partial E$ of the carrier momentum relaxation time  $\tau_{\mu}$.  This contrasts with the resistivity $\rho$ which depends only on $\tau_{\mu}$ itself.  Furthermore, as first shown by Obraztsov \cite{Obraztsov}, the thermopower of a non-interacting electron gas is closely related to the \emph{entropy} per particle in the system.  This remarkable connection between a transport coefficient and a fundamental thermodynamic quantity was later shown to hold even for strongly interacting electrons at high magnetic field, at least in the clean limit \cite{Cooper}.  This is strong motivation for the study of thermopower since the entropy of certain fractionally quantized Hall phases (notably the even-denominator state at Landau level filling factor $\nu = 5/2$) may be anomalously large if the relevant quasiparticles exhibit non-abelian exchange statistics \cite{MooreRead,Yang}.

The outline of the paper is as follows.  Section II describes our heterostructure sample and the experimental method by which we determine the longitudinal thermopower $S$ of the 2DES within it.  For ease of reading, some of the technical details have been relegated to the Appendix.  In section III we describe and discuss our results.  At zero magnetic field we demonstrate that the thermopower in our sample is dominated by electron diffusion at temperatures below about $T = 150$ mK, and is well described by the famous Mott formula \cite{Cutler}.  At high magnetic fields we report results for both the compressible composite fermion fluid at Landau level filling factor $\nu = 3/2$ and the incompressible fractional quantized Hall state at $\nu = 5/2$. At $\nu = 3/2$ we again observe a diffusive thermopower at low temperatures.   We extract an estimate of the composite fermion effective mass in this compressible state and compare it to both theory and prior experiments.  At $\nu = 5/2$ our data clearly reveal the incompressibility of this exotic collective state and thereby allow us to discuss our results within the context of a recent theoretical model.

\section{\label{sec:level1}Experimental Method}

The basic technique we employ is standard: A temperature gradient is imposed along the length of a bar-shaped sample containing the 2DES by applying heat to one end while the other end is thermally grounded. Thermoelectric voltages occurring within the 2DES in the bar are then recorded as functions of magnetic field and average sample temperature. In our one departure from common practice, no external thermometers are mounted on the sample. Instead, the resistivity of the 2DES, which is partitioned into two distinct regions, is employed to provide thermometry. The details of this method are presented below.

\subsection{\label{sec:level2}Sample}

The sample used in this experiment is a GaAs/AlGaAs heterostructure grown by molecular beam epitaxy on a $\langle001\rangle$-oriented GaAs substrate. The crucial epilayers in the sample consist of a 30 nm GaAs quantum well flanked by thick Al$_{0.24}$Ga$_{0.76}$As layers. Symmetrically placed Si doping sheets in the Al$_{0.24}$Ga$_{0.76}$As layers create a 2DES in the lowest subband of the GaAs quantum well. After illumination by a red LED, the density and mobility of the 2DES are $N = 2.9\times 10^{11}$ cm$^{-2}$ and $\mu = 3.1\times 10^{7}$ cm$^{2}$/Vs at low temperatures. The sample exhibits a wealth of fractional quantized Hall effect (FQHE) states, including a robust $\nu = 5/2$ state as shown in Fig. 1. The temperature dependence of the longitudinal resistance $R_{xx}$ at $\nu = 5/2$ reveals a charge gap $\Delta_{5/2} \approx 450$ mK. Note that the data in Fig. 1, the measured charge gap $\Delta_{5/2}$, and all of the thermopower data discussed in this paper were obtained from the same sample.

\begin{figure}
\includegraphics[width=3.2in, bb=10 0 280 192]{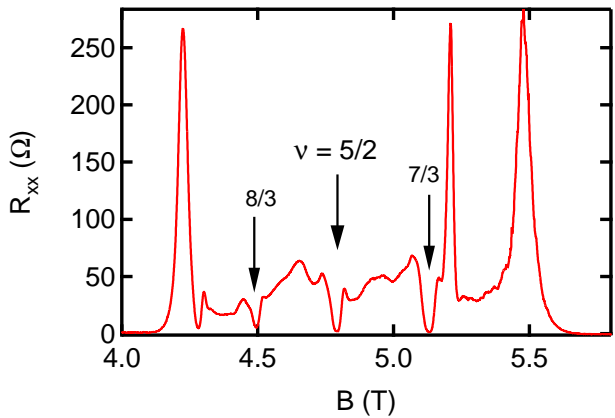}
\caption{\label{}Longitudinal resistance $R_{xx}$ {\it vs.} magnetic field at $T = 50$ mK in the first excited Landau level.  Minima in $R_{xx}$ at the $\nu = 7/3$, 5/2, and 8/3 FQHE states are indicated.} 
\end{figure}
The sample is bar-shaped, $L = 12$ mm long by $W =3$ mm wide. A strain gauge heater \cite{StrainGauge} is attached to one end of the sample while the other end is indium soldered to a small Au-plated copper slab that serves as thermal ground. The copper slab is in turn bolted to the cold finger of a dilution refrigerator. A schematic diagram of the sample is shown in Fig. 2. In order to reduce the phonon mean free path in the sample (and thereby the influence of phonon drag on the thermopower), the substrate is thinned to about $t = 130$ $\mu$m via a chemical etch and then sandblasted to ensure diffuse phonon scattering \cite{Klitsner}. This latter step is important, for otherwise highly specular phonon boundary scattering can lead to non-uniform temperature gradients along the bar.

The front surface of the sample is also etched, leaving the 2DES intact only in two square 3 by 3 mm mesas, separated by 1 mm and positioned symmetrically about the midpoint of the bar. Transport and thermoelectric measurements on each 2DES mesa are enabled by six small In-Sn ohmic contacts diffused onto its perimeter. As the diagram suggests, the two mesas share one ohmic contact and are thereby connected in series.  Manganin wires, 25 $\mu$m in diameter and approximately 1 cm long, are attached to these contacts (and to the resistive heater). The thermal conductance and thermopower \cite{manganin} of these wires is negligible in comparison to the thermal conductance of the sample bar and the thermopower of the 2D electron systems within it.    


\subsection{\label{sec:level2}Thermal Conductance Calibration}

Before thermoelectric measurements can be performed, the thermal conductance of the sample bar must be determined. This conductance is overwhelmingly dominated by phonon transport; diffusive heat transport by the 2DES is negligible. Once the thermal conductance is established, a known temperature gradient can be imposed and the resulting thermoelectric voltages in the 2DESs can be converted into measurements of thermopower.

Our procedure for determining the thermal conductance $K$ of the sample bar is as follows. A magnetic field applied perpendicular to the sample establishes the quantum Hall effect regime in the 2DESs. The field is chosen to be in the vicinity of an integer quantized Hall state (e.g. at $\nu = 1$, 2, or 3) where the longitudinal resistance $R_{xx}$ of the 2DESs is strongly temperature dependent. Then, in response to a small step change $\Delta T_0$ in the cold finger temperature, the resistance changes $\Delta R_{xx,1}$ and $\Delta R_{xx,2}$ of each 2DES are recorded. Since this is done without applying any power to the strain gauge heater on the sample, the temperatures $T_{1,2}$, and temperature changes $\Delta T_{1,2}$, of the two 2DES regions are assumed to be the same and equal to those of the cold finger, $T_0$ and $\Delta T_0$. Next, with the cold finger temperature held constant at $T_0$, a small heat flux $Q$ is applied to the strain gauge heater \cite{Q}. $Q$ is chosen to render the resulting resistance changes $\Delta R_{xx,1}$ and $\Delta R_{xx,2}$ of the two 2DES regions comparable to those observed when the cold finger temperature was changed in the previous measurement. Comparison of these resistance changes with those observed when the cold finger temperature was changed allows the temperature rises $\Delta T_{1,2}$ to be determined. As expected, $\Delta T_1 < \Delta T_2$ since there is now a temperature gradient along the sample bar and mesa 2 is farther from thermal ground than is mesa 1. The thermal conductances $K_{1,2}$, between each 2DES region and thermal ground, are then given by $K_1 = Q/\Delta T_1$ and $K_2 = Q/\Delta T_2$. Note that $\Delta T_{1,2}$ is kept $\leq 10\%$ of $T_0$ during these thermal conductance measurements.

Importantly, we find that $K_2/K_1 = L_1/L_2 = 0.49$ to within experimental uncertainty, with $L_1 = 3.7$ mm and $L_2 = 7.6$ mm the distances between the midpoint of the respective 2DES mesa and the indium solder joint that connects the sample to the cold finger. This simple geometric scaling proves that the thermal resistance of the solder joint itself is negligible in comparison to that of the sample. 

\begin{figure}
\includegraphics[width=3.0in, bb=10 0 275 470]{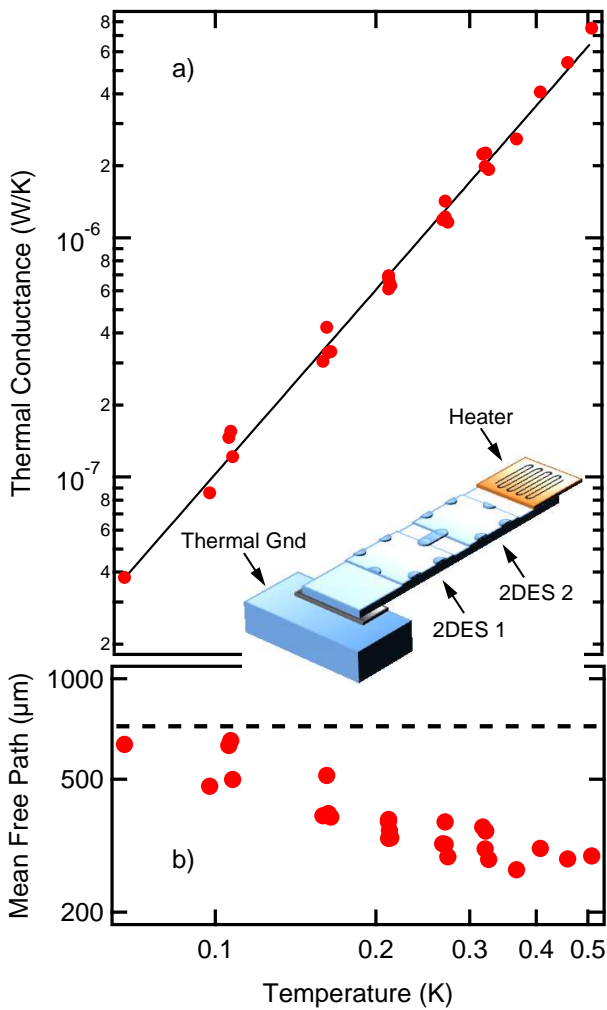}
\caption{\label{}(color online) a) Measured thermal conductance $K_2$ of between mesa 2 and thermal ground {\it vs.} temperature. The solid black line is a fit to the data and scales as $T^{2.56 \pm 0.05}$. b) Phonon mean free path inferred from the thermal conductance data. The dashed line corresponds to 720 $\mu$m, an estimate of the diffuse boundary scattering limit. Inset: Device schematic. Note that the two 2DES mesas share one ohmic contact.}
\end{figure}
Figure 2(a) displays the thermal conductance $K_2$ between 2DES mesa 2 and thermal ground as a function of temperature $T$ in a log-log plot.  The figure demonstrates that $K_2$ follows a simple power law temperature dependence: $K_2 \propto T^{2.56 \pm 0.05}$. According to simple gas kinetic theory, the low temperature phonon thermal conductance is $K_2 = \gamma Cv_{ph}\Lambda /3$, with $C$ the lattice specific heat of GaAs\cite{Cetas}, $v_{ph} = 3300$ m/s the appropriate mean acoustic phonon velocity \cite{velocity}, $\Lambda$ the phonon mean free path, and $\gamma = Wt/L_2$ the cross-sectional area to length ratio of the bar. Since $C \propto T^3$ at these low temperatures, our data demonstrate that the phonon mean free path scales roughly as $\Lambda \propto T^{-0.44}$ over the temperature range studied here. Figure 2(b) shows the deduced values of $\Lambda$ in $\mu$m. As the temperature is reduced, the mean free path grows and approaches the diffuse boundary scattering limit, $\Lambda_b$, estimated to be $\approx 720$ $\mu$m (the dashed line in Fig. 2(b)). This estimate is based on the assumption that diffuse scattering occurs only at the substrate side of the sample. That $\Lambda$ is considerably larger than the sample thickness is a result of the high aspect ratio of the sample cross-section ($W/t \approx 23$) \cite{Wybourne}.

The thermal conductance data of Fig. 2(a) were acquired at a variety of magnetic fields in the vicinities of the QHE states at $\nu$ = 1, 2, and 3. While minor systematic variations were found (and are evident in the data scatter in Fig. 2), no unambiguous magnetic field dependence emerged.  This is not surprising since the thermal conductivity of our sample is heavily dominated by phonon transport. Nevertheless, the resulting uncertainty in $K_2 (T)$ is an important source of systematic error in the present thermopower experiment. We estimate that the uncertainty in $K_2$ translates into a relative uncertainty of about 7\% in thermopower.

\subsection{\label{sec:level2}Thermoelectric Measurements}

Once the thermal conductance of the sample bar is known, we can perform thermoelectric measurements, translating thermovoltages into thermopower. Applying a heat flux to the strain gauge gives rise to a temperature gradient along the length of the sample. The voltage along a 2DES mesa is then measured using a low noise DC amplifier \cite{EM}. From the thermal conductance data of Fig. 2(a), along with the known cold finger temperature $T_0$ and the applied heat flux $Q$, we calculate the temperature difference $\Delta T$ between the ohmic contacts used to measure the thermovoltage as well as the mean temperature $T$ of the 2DES to which the measurement applies. The measured longitudinal thermopower, or Seebeck coefficient, is then given by $S(T) \equiv -\Delta V/\Delta T$, where $\Delta V$ is the thermovoltage measured along the 2DES mesa.

For much of the data reported here the time required for the 2DES to relax to steady state following switching the heater on or off is short enough that a conventional lock-in technique may be used to measure the thermoelectric voltages. However, at the lowest temperatures and highest magnetic fields, extremely long thermal relaxation times (several minutes) are encountered.  The temperature and magnetic field dependence of these relaxation times suggests that nuclear moments, most likely in the In-Sn ohmic contacts and the manganin wires attached to them, are responsible.  To contend with this issue, we employ a quasi-DC data acquisition and analysis technique that allows measurements at very low frequencies (e.g. 1 mHz) and the easy rejection of data acquired before the sample has reached steady state. A more detailed discussion of these issues is given in the Appendix. 

Figure 3 illustrates our thermopower measurement protocol with data taken at filling factor $\nu = 3/2$.  The figure shows the time dependence of the observed longitudinal voltage drop $V_{DC}$ between ohmic contacts on 2DES mesa 2 as the heater power is toggled between $Q = 0$ and 18 nW.  The deduced thermoelectric voltage $\Delta V$ is taken to be the average difference between $V_{DC}$ with the heat on versus off; for the present example this is $\Delta V \approx 78$ nV.  For these data the cold finger temperature was maintained at $T_0 = 120$ mK.  Integrating the thermal conductance data of Fig. 2(a) reveals that for $Q = 18$ nW, 2DES mesa 2 is at a mean temperature of $T_2 \approx 180$ mK with a temperature difference of $\Delta T \approx 14$ mK between the ohmic contacts. Combining these numbers yields a thermopower of $S = -\Delta V/\Delta T \approx -5.6$ $\mu$V/K.

In the above example, the thermoelectric voltage $V_{DC}$ is measured along 2DES mesa 2 (the one farthest from thermal ground) using the ohmic contacts that lie on the central axis of the sample bar (see the inset to Fig. 2). This is the case for all the measurements reported here except those done at zero magnetic field. In that case the net voltage difference across both 2DES mesas is recorded, with the one ohmic contact they share providing the on-chip series connection.  This was done in order to reduce the relative uncertainty in the distance between the ohmic contacts.  This procedure was not applied at high fields owing to the slight 2DES density differences between the two mesas ($\Delta N/N \sim 1\%$) which, while small, can lead to differences in Landau level filling fraction that are comparable to the width of important fractional quantized Hall states (notably at $\nu = 5/2$). 

\begin{figure}
\includegraphics[width=3.0in, bb=20 0 265 218]{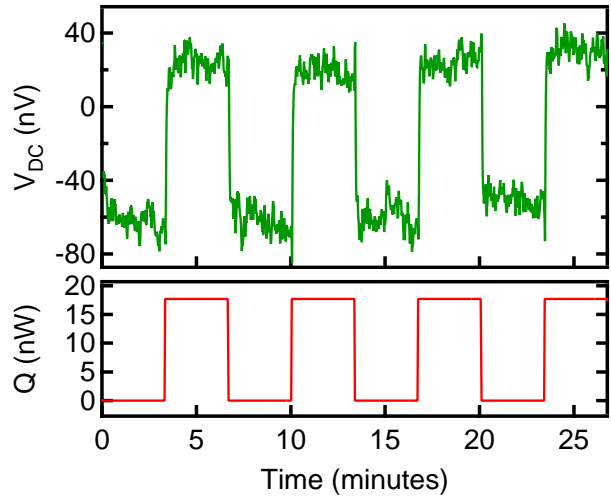}
\caption{\label{} Upper panel: Voltage $V_{DC}$ measured across 2DES mesa 2 {\it vs.} time as heater is toggled on and off. Lower panel: Power $Q$ applied to strain gauge heater {\it vs.} time. In this example, taken at $\nu = 3/2$, the cold finger is at 120 mK such that 2DES mesa 2 is at 180 mK with $\Delta T$ = 14 mK across the mesa when heat is applied.} 
\end{figure}
Finally, we comment on the non-zero voltages which are observed, as Fig. 3 reveals, even when the heater is off.  We attribute these voltages to offsets and/or $1/f$ noise in our DC amplifier as well to genuine thermoelectric effects arising from a lack of perfect thermal symmetry in the measurement circuit.  In any case, such background voltages, which vary slowly with time, are unrelated to thermoelectric phenomena in the 2DES and are readily subtracted.

\section{\label{sec:level1}Results and Discussion}

\subsection{Zero Magnetic Field}
Figure 4 displays our results for the thermopower of 2D electrons at zero magnetic field.  As the dashed line suggests, the observed thermopower is proportional to temperature below about 150 mK.  This is the expected result when electron diffusion dominates the thermopower.  It has long been known that both diffusion and phonon drag \cite{Gallagher, Fletcher2} contribute to the thermopower of 2D electrons in GaAs/AlGaAs heterostructures.  Not surprisingly, the phonon contribution subsides rapidly as the temperature falls, leaving the diffusion contribution dominant. The precise cross-over temperature is non-universal, depending on various extrinsic factors, notably the phonon mean free path. As Fig. 4 demonstrates, in our sample the thermopower begins to exceed the extrapolated linear temperature dependence due to diffusion at about $T = 200$ mK.  A similar cross-over temperature was observed by Ying {\it et al.} \cite{Ying} in their study of the thermopower of 2D {\it hole} systems in GaAs. We note in passing that it is possible to dramatically reduce the phonon drag contribution using a hot-electron technique developed recently by us \cite{hot_electron}, but this method has not yet been applied to the high magnetic field regime which is the main focus of the present work.  

\begin{figure}
\includegraphics[width=3.2in, bb=10 0 285 231]{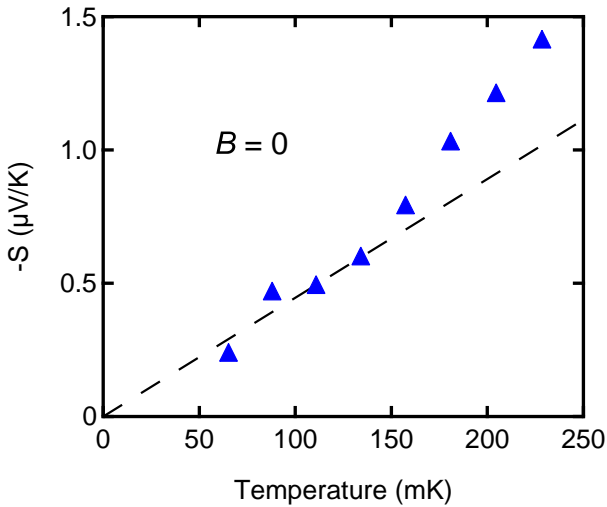}
\caption{\label{}Zero field thermopower {\it vs.} temperature. The dashed line represents the thermopower of Eq. (1) for $\alpha = 0.9$.} 
\end{figure}

According to the Mott formula \cite{Cutler}, at zero magnetic field the diffusion thermopower $S^d$ of a 2DES that behaves as a simple Drude metal is, in the $T\rightarrow0$ limit,

\begin{equation}\label{eq.1}
S^d =-\frac{\pi k_B^2}{3e}\frac{m^*}{N \hbar^2}(1+\alpha)T
\end{equation}
where $m^*$ is the band mass of the electrons ($m^*=0.067m_e$ in GaAs), $N$ is the 2DES density ($N = 2.9 \times 10^{11}$ cm$^{-2}$ in our sample) and $\alpha$ reflects the energy and thus density dependence of the electronic momentum scattering time $\tau$: $\alpha \equiv(N/\tau)\frac{d\tau}{dN}$. The parameter $\alpha$ depends on the details of the dominant electron scattering mechanism in the sample and is typically $0.5 \lesssim \alpha \lesssim 1.5$ for 2D electrons in modulation-doped GaAs heterostructures \cite{Hwang}.  Fitting Eq. 1 to our $T<150$ mK data yields $\alpha \approx 0.9$ in our sample; this fit is shown as a dashed line in Fig. 4.  Since the 2D density in our sample is not adjustable via electrostatic gating \cite{no_gating}, it is not possible to independently determine $\alpha$ from density-dependent resistivity measurements. Nevertheless, the good agreement between our low temperature data and the Mott formula (with a reasonable value of $\alpha$) gives us confidence in the reliability of our experimental technique for measuring thermopower.

\subsection{Composite Fermion Metal at $\nu = 3/2$}
Figure 5 shows the measured thermopower {\it vs.} temperature at $B = 8$ T where the Landau level filling fraction is $\nu = 3/2$.  As the figure demonstrates,  below about $T = 200$ mK the thermopower at $\nu = 3/2$ is simply proportional to the temperature, $T$.  At higher temperatures this proportionality is lost, presumably due to the increasing importance of phonon drag \cite{Ying,Bayot,Ruf}.  Qualitatively, these $\nu = 3/2$ results are very similar to those observed at $B = 0$; only the magnitudes are different.  In the low temperature regime the thermopower at $\nu = 3/2$ is roughly 7 times larger than at $B = 0$.

At $\nu = 3/2$ the 2DES is in a compressible, i.e. gapless state.  Conjugate to the $\nu = 1/2$ state in the lower spin branch of the lowest Landau level, the $\nu =3/2$ state is best understood within the Chern-Simons composite fermion (CF) theory \cite{HLR}.  Crudely speaking, at both of these filling factors the 2DES may be viewed as a Fermi liquid of weakly interacting CFs in {\it zero} effective magnetic field.  Exploiting the connection \cite{ Obraztsov} between thermopower and entropy per particle, Cooper {\it et al.} \cite{Cooper}
arrive at a Mott-like formula for the diffusion thermopower $S_{CF}^d$ of a disorder-free 2DES at $\nu =$ 1/2 and 3/2:

\begin{equation}\label{eq.2}
S_{CF}^d =-\frac{\pi k_B^2}{6e}\frac{m_{CF}^*}{N \hbar^2}T
\end{equation}
where $N$ is the total 2DES density and  $m_{CF}^*$ is the effective mass of the CFs.  This remarkably simple result differs from Eq. 1 in three ways.  First, Eq. 2 assumes that the CF spins are fully polarized, rather than completely unpolarized as in Eq. 1.  This accounts for the factor of 6 in the denominator of Eq. 2 replacing the factor of 3 in Eq. 1.  Obviously, whether the spins are in fact fully polarized at $B = 8$ T and $\nu = 3/2$ in our sample can be questioned.  Second, since Eq. 2 applies to an idealized disorder-free 2DES, the $\alpha$ parameter in Eq. 1 does not enter.  However, Cooper {\it et al.} \cite{Cooper} note that impurity scattering of CFs is in any case expected to be only weakly energy dependent, with $\alpha = 0.13$ being one theoretical estimate \cite{Khveshchenko}.  Finally, there is the substantial issue of the effective mass $m_{CF}^*$ of the CFs replacing the GaAs $\Gamma$-point conduction band mass $m^* = 0.067 m_e$ (with $m_e$ the bare electron mass) in the $B = 0$ formula.  The CF mass, which derives entirely from electron-electron interactions, is generally much larger than the band mass and this, we believe, is the main reason for the much larger thermopower at $\nu = 3/2$ compared to $B = 0$.

\begin{figure}
\includegraphics[width=3.2in, bb=20 0 280 231]{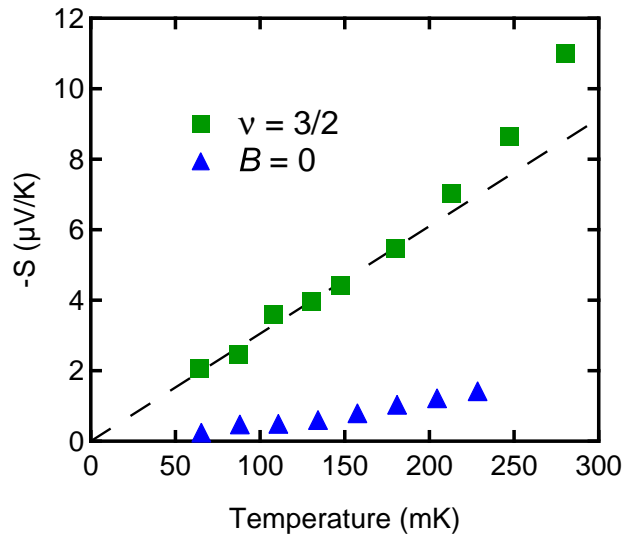}
\caption{\label{}(color online) Thermopower {\it vs.} temperature at $\nu = 3/2$ (green squares) along with that at zero field (blue circles).  The dashed line represents the thermopower of Eq. (2) for $m_{CF}^* \approx 1.7m_e$.} 
\end{figure}

As the dashed line in Fig. 5 shows, a CF effective mass of $m_{CF}^*=1.7m_e$ provides an excellent fit of Eq. 2 to our $\nu = 3/2$, $B = 8$ T thermopower data below 200 mK.  This value of the effective mass is roughly 7 times larger than the predicted \cite{HLR,Morf} CF mass at the Fermi surface: $ m_{CF}^* \approx 0.085\sqrt{B[T]}m_e = 0.24 m_e$ at $B=8$ T. Interestingly, but perhaps coincidentally, our measured value of $m_{CF}^*$ instead agrees well with a recent theoretical estimate of the so-called polarization mass \cite{Park}. However, it seems to us unlikely that the polarization mass, which depends on the full depth of the Fermi sea rather than just its surface properties, plays an important role in the thermal transport properties of the system.

While origin of the discrepancy between theory and experiment on the $\nu = 3/2$ effective mass is not known, there are several points to consider.  On the theoretical side, the above estimates only apply to infinitely thin, disorder-free 2D systems at magnetic fields large enough that Landau level mixing is negligible.  In general, violating any of these approximations tends to reduce the energy gap for the fractional quantized Hall states.  Since in the CF model these gaps are inversely proportional to $ m_{CF}^*$, inclusion of these non-idealities would increase the theoretical estimates of the mass and thereby reduce the disagreement between theory and experiment.  On the experimental side, the spin polarization and the value of $\alpha$ (which reflects the energy dependence of the CF scattering rate) at $\nu = 3/2$ are obvious sources of uncertainty.  For example, if the spins were completely depolarized and $\alpha \sim 1$ as at zero magnetic field, then the CF effective mass deduced from our $\nu = 3/2$ thermopower data would be reduced by a factor of 4.  However, this scenario seems unlikely since spin polarization experiments \cite{Kukushkin} at $\nu = 3/2$ indicate close to maximum polarization at $B = 8$ T and theoretical estimates \cite{Khveshchenko} of $\alpha$ are small ($\alpha \approx 0.13$).  

Similar thermopower experiments have been performed previously on 2D {\it hole} systems \cite{Ying}, albeit in a much lower density sample than ours.  Following Cooper {\it et al.} \cite{Cooper} we deduce from the data of Ying {\it et al.} \cite{Ying} a CF effective mass of $m_{CF}^* \approx 1.3 m_e$ at $B = 5.6$ T by applying Eq. 2 (which assumes complete spin polarization and $\alpha = 0$).  Since $m_{CF}^*$ should scale as $\sqrt{B}$, this would imply $m_{CF}^* \approx 1.6 m_e$ at $B = 8$ T, close to the value we deduce from our $\nu =3/2$ data at that magnetic field.  Interestingly, at $\nu = 3/2$ and $B = 1.87$ T Ying {\it et al.} find a thermopower roughly 5 times larger than we do at the same filling factor, but at $B = 8$ T.  After accounting for the difference in density between the two samples, these two results cannot be reconciled using Eq. 2 and the assumption that $m_{CF}^* \propto \sqrt{B}$. However, if we make the further assumption that in the Ying {\it et al.} sample the $\nu = 3/2$ spin polarization is near zero while in our much higher density sample the polarization is maximal, then consistency can be obtained.  The large difference in the $\nu = 3/2$ magnetic fields of the two samples makes this a plausible, if unproven, assumption.

Finally, it is worth pausing to consider the physical origin of the linear temperature dependence of the thermopower at $\nu =$ 1/2 and 3/2.  As noted by Cooper {\it et al.} \cite{Cooper}, this follows from the linear temperature dependence of the entropy of a Fermi liquid of CFs.  However, unlike the situation at zero magnetic field, the Fermi liquid at $\nu =$ 1/2 and 3/2 only exists because of interactions between electrons.  Indeed, if interactions (and disorder) are ignored, the entropy per electron of a partially filled Landau level is determined solely by trivial combinatorics.  At $\nu = 3/2$ the implied thermopower would then be $S_{3/2}^d = 2 k_B$ln(2)/3$e \approx 40 \mu$V/K, {\it independent of temperature} as $T \rightarrow 0$.  The much smaller, and linearly temperature dependent thermopower that we observe at $\nu = 3/2$ is thus a dramatic signature of the entropy-reducing, or ``organizing'' effects of Coulomb interactions. 

\subsection{FQHE State at $\nu = 5/2$}
We turn now to the incompressible fractional quantized Hall state at filling factor $\nu = 5/2$, ignoring for the moment the intriguing subtleties of this state alluded to in the Introduction. Unlike the compressible states at $B =0$ and at $\nu = 3/2$, the presence of an energy gap at the Fermi level will strongly suppress the entropy at low temperatures and prevent a linear temperature dependence of the thermopower $S_{5/2}^d$ from appearing at $\nu = 5/2$. Instead, one anticipates that $S_{5/2}^d$ will vanish much more rapidly with falling temperature, possibly in a thermally activated manner. Behavior of this kind has previously been qualitatively observed in thermopower measurements at other quantized Hall states, both integer and fractional \cite{Bayot,Ruf}. 

\begin{figure}
\includegraphics[width=3.2in, bb=10 0 270 232]{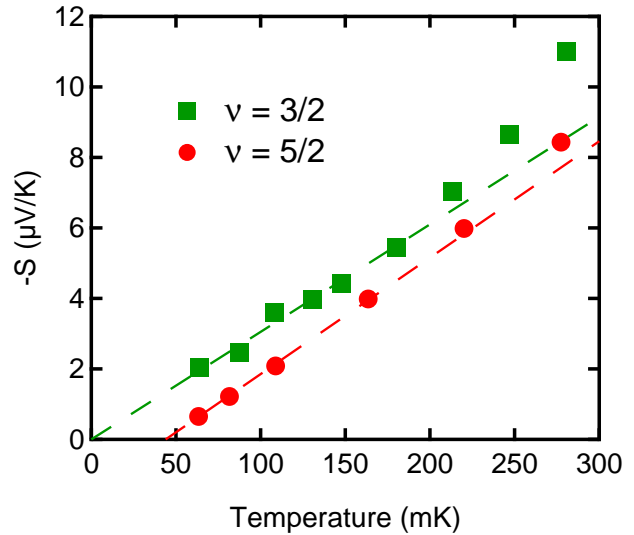}
\caption{\label{}(color online) Comparison of measured thermopowers {\it vs.} temperature at $\nu = 5/2$ (red circles) and $\nu = 3/2$ (green squares). Unlike the thermopower at $\nu = 3/2$, the thermopower data at $\nu = 5/2$ extrapolate to zero at a {\it non-zero} temperature.} 
\end{figure}

Figure 6 compares the temperature dependence of the thermopower at $\nu = 5/2$ with that at $\nu = 3/2$.  Unlike the $\nu =3/2$ data, the $S_{5/2}^d$ data cannot be well fit by a straight line passing through the origin. Instead, a linear fit to the $T<250$ mK data, indicated by the red dashed line in the figure, extrapolates to $S_{5/2}^d = 0$ at about $T = 44$ mK.  We believe this is a reflection of the energy gap in the $\nu = 5/2$ FQHE state.

\begin{figure}
\includegraphics[width=3.2in, bb=0 0 300 232]{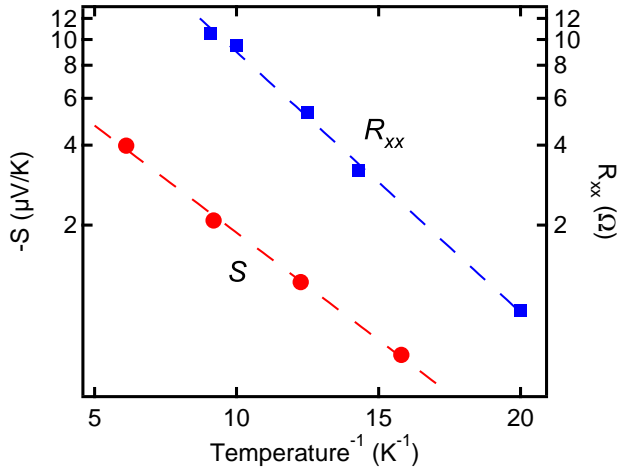}
\caption{\label{}(color online) Arrhenius plots of thermopower $S$ (red circles, left axis) and longitudinal resistance $R_{xx}$ (blue squares, right axis) at $\nu = 5/2$.  Dashed line fits give energy gaps of $\Delta \approx 370$ mK and 450 mK for the thermopower and resistance, respectively.} 
\end{figure}
Figures 7 and 8 support our assertion that the low temperature thermopower at $\nu = 5/2$ is dominated by the FQHE energy gap.  Figure 7 compares, in an Arrhenius plot, the temperature dependence of the thermopower $S_{5/2}^d$ and the longitudinal resistance $R_{xx}$ (both measured in mesa 2) at $\nu = 5/2$.  In spite of the somewhat limited data set, it is clear from the figure that both the thermopower and the longitudinal resistance are consistent with simple thermal activation ($i.e.$ both scale as $\sim e^{-\Delta/2T}$) for the roughly one order of magnitude variation of each data set. From the slopes of the dashed lines in the figure, we find $\Delta \approx 370$ mK and 450 mK for the thermopower and resistivity data, respectively. These values are quite comparable to those obtained from previous resistivity measurements at $\nu = 5/2$ in 2DES samples of similarly high quality \cite{jpe2002,pan,dean}. 

\begin{figure}
\includegraphics[width=3.2in, bb=5 0 301 232]{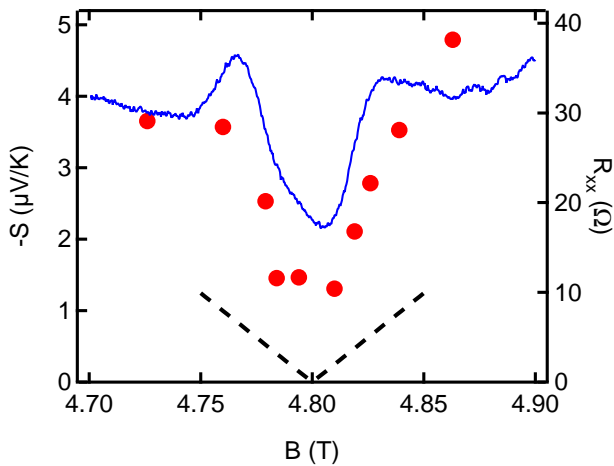}
\caption{\label{}(color online) Thermopower {\it vs.} magnetic field (red circles) along with $R_{xx}$ {\it vs.} magnetic field (blue curve) about $\nu = 5/2$ at $T \approx 82$ mK. The dashed line represents the thermopower of Eq. (3) for $B_0 = 4.80$ T.} 
\end{figure}

Figure 8 demonstrates that the low temperature ($T \approx 82$ mK) diffusion thermopower of the 2DES exhibits a clear minimum versus magnetic field at $\nu = 5/2$. The magnetic field location of the thermopower minimum coincides, within experimental uncertainty, with that of the longitudinal resistance $R_{xx}$, which is also shown in the figure.  Taken together, Figs. 7 and 8 convincingly demonstrate that the incompressibility of the $\nu = 5/2$ FQHE state is detectable in the thermopower of the 2D electron system.   

The incompressible ground state of a 2DES at $\nu = 5/2$ is currently believed to be well approximated by the Moore-Read, or Pfaffian, wavefunction \cite{MooreRead}.  This state, which may be viewed as a BCS condensate of p-wave paired composite fermions, has come under intense scrutiny recently owing to the expected non-abelian exchange statistics of its quasiparticle excitations.  Unlike conventional abelian FQHE states (e.g. at $\nu = 1/3$), multiple pairwise interchanges (braidings) of localized non-abelian quasiparticles at $\nu = 5/2$ generate a large Hilbert space of degenerate ground states.  This Hilbert space, which is topologically protected from local disturbances that might otherwise lead to decoherence, has been suggested as an ideal venue for the storage and processing of quantum information \cite{dassarmaRMP}.

The anomalous ground state degeneracy arising from non-abelian quasiparticle statistics is anticipated to have observable consequences in certain thermal transport and thermodynamic measurements \cite{Nayak,Yang,CooperStern}. Most relevant here is the prediction \cite{Yang} of Yang and Halperin (YH) that the excess entropy arising from the ground state degeneracy of a collection of localized non-abelian quasiparticles is expected, under certain conditions, to dominate the thermopower of the 2DES near filling factor $\nu = 5/2$.  YH find that under ideal circumstances the low (but not too low \cite{Yang}) temperature thermopower of a 2DES very near $\nu =  5/2$ is {\it temperature independent} and proportional to $|1 –- B/B_0|$, where $B_0$ is the magnetic field corresponding to $\nu = 5/2$. (A deviation of the magnetic field from $B_0$ is necessary, in the ideal case, to produce quasiparticles in the ground state of the system.  In real samples quasiparticles are doubtless present even at $B = B_0$ and $T = 0$ owing to density inhomogeneities and other forms of disorder.)  

The dashed lines in Fig. 8 represent the quantitative prediction \cite{Yang} of YH for the thermopower of the 2DES near $\nu  = 5/2$.  Clearly, this prediction underestimates the experimentally observed thermopower. While it is not yet possible to unambiguously identify the sources of the excess thermopower we observe, certain possibilities come to mind.  First, as YH stress, their calculation is for an ideal, disorder-free 2DES.  Our sample, while extraordinarily pure, is certainly not disorder-free.  Indeed, the very existence of Hall plateaus demonstrates this.  Second, the YH prediction applies only within a somewhat difficult to quantify temperature window.  In particular, the relatively high temperature ($T \approx 82$ mK) to which the data in Fig. 8 pertain may be sufficient to activate other sources of entropy within the 2DES.  

While the present results do not support (or refute) the existence of excess entropy at $\nu = 5/2$ arising from non-abelian quasiparticle exchange statistics, there is reason to hope that future thermopower experiments may do so.  The results in Fig. 8 show that the gap between theory and experiment is not enormous.  In particular, if the very long thermal relaxation times encountered here can be overcome, then lower temperature, higher resolution data can be obtained.  

\section{\label{sec:level1}Conclusion}

We have measured the longitudinal thermopower $S$ of ultra-high mobility two-dimensional electrons at both zero magnetic field and at high fields in the compressible metallic state at filling factor $\nu = 3/2$ and the incompressible fractional quantized Hall state at $\nu=5/2$.

At zero magnetic field, we find that $S$ is dominated by electron diffusion below about $T =150$ mK. From the linear temperature dependence in this diffusion dominated regime, we estimate the parameter $\alpha$, which reflects the energy dependence of the momentum scattering time, to be $\alpha \approx 0.9$ in our sample. Above about 200 mK, phonon drag begins to contribute significantly to the thermopower and $S$ becomes superlinear in $T$. 

At high magnetic field, in the compressible state at filling factor $\nu = 3/2$, $S$ exhibits a temperature dependence very similar to that at $B = 0$, only the magnitude of the thermopower is about 7 times larger. Below about $T = 200$ mK $S$ is linear in temperature and comparison to a recent theory of thermopower of composite fermions \cite{Cooper} allows us to estimate the CF mass to be $m_{CF}^* \approx 1.7 m_e$, with $m_e$ the bare electron mass. This estimate of $m_{CF}^*$ substantially exceeds theoretical estimates of the density of states mass of spin polarized composite fermions \cite{HLR,Morf}.  

Unlike the compressible states at $B = 0$ and $\nu = 3/2$, at $\nu = 5/2$ the temperature dependence of $S$ at low temperatures suggests the influence of the FQHE energy gap $\Delta$.  The presence of a clear minimum in the magnetic field dependence of $S$ centered at $\nu = 5/2$ confirms that the incompressibility of the 5/2 state is readily detectable in the electronic thermopower. From our $S$ measurements we estimate $\Delta \approx 370$ mK, which compares well with the value $\Delta \approx 450$ mK obtained from resistivity measurements on the same sample. 

The observed magnitude of $S$ {\it vs.} $B$ around $\nu = 5/2$ exceeds recent estimates \cite{Yang} which take account of the anticipated non-abelian exchange statistics of the quasiparticles of this intriguing fractional quantized Hall state.  While the sources of the excess entropy influencing our results are as yet unknown, the discrepancy between theory and experiment is relatively modest and leaves us optimistic that future experiments, at lower temperatures and in cleaner samples, may yet require non-abelian quasiparticles for their quantitative understanding. 

We thank Nigel Cooper, Bert Halperin, Jainendra Jain, Gil Refael, Steve Simon, Ady Stern, and Kun Yang for helpful discussions. This work was supported via DOE grant DE-FG03-99ER45766 and Microsoft Project Q.

\section{\label{sec:level1}Appendix}

In this section we address two important technical aspects of our experiment:  The long thermal relaxation times observed at low temperatures and high magnetic fields, and the quasi-dc data acquisition and analysis methods we applied to successfully measure the thermopower in the presence of them.  

\subsection{Thermal Relaxation Times}
At the lowest temperatures, particularly when a large magnetic field is applied, the time required for the sample to achieve steady state following the turning on or off the heater becomes very long. These long time constants have forced us to acquire data at extremely low frequencies ($\sim 1$ mHz). 

The thermal time constant of our sample is readily observed via measurements of the longitudinal resistance $R_{xx}$ of the 2DES.  After choosing a magnetic field where $R_{xx}$ is strongly temperature dependent, its time evolution following an abrupt change in the heater power is recorded. The inset to Fig. 9 shows a typical example: After turning off the heater at $T \approx 75$ mK and $B = 6.4$ T, $R_{xx}$ takes over 100 seconds to fully relax.  Fitting such relaxations to a simple exponential, $\Delta R_{xx} \propto e^{-t/\tau_R}$, allows us to extract the relaxation time $\tau_R$.  Figure 9 shows $\tau_R$ {\it vs.} temperature at various magnetic fields in a log-log plot.  We find the temperature dependence of $\tau_R$ is reasonably approximated by a simple power law, $\tau_R \propto T^p$, with the exponent $p \approx -3.7$ essentially independent of magnetic field.

To more clearly illustrate the magnetic field dependence of $\tau_R$, Fig. 10 displays $\tau_R$ {\it vs.} $B$ in a log-log plot. It is obvious that $\tau_R$ {\it vs.} $B$ does not obey a simple power law.  Indeed, $\tau_R$ appears to saturate at a strongly temperature dependent value in the $B \rightarrow 0$ limit.  At high fields $\tau_R$ becomes strongly field dependent, becoming roughly consistent with $\tau_R \propto B^2$ beyond $B \sim 6$ T.   

\begin{figure}
\includegraphics[width=3.0in, bb=30 25 300 265]{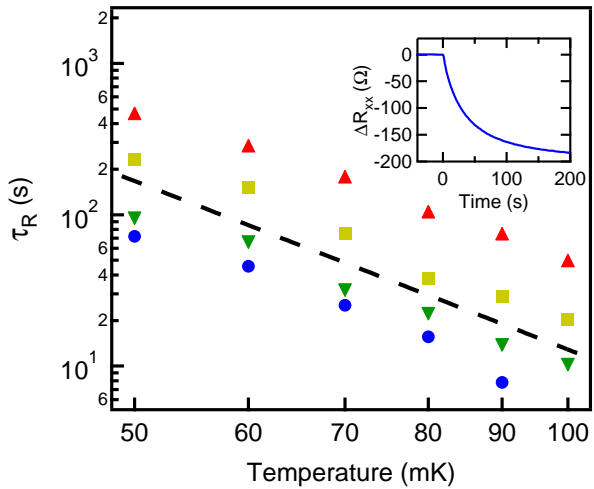}
\caption{\label{}(color online) Thermal relaxation time $\tau_R$ {\it vs.} temperature at various magnetic fields. From top to bottom: $B$ = 10.2, 6.4, 3.0, and 1.2 T. Dashed line shows a simple power law: $\tau_R \propto T^{-3.7}$. Inset: Change in the longitudinal resistance, $\Delta R_{xx}$, {\it vs.} time after turning off the heater (in this example, at $T \approx 75$ mK and $B = 6.4$ T).}
\end{figure}
\begin{figure}
\includegraphics[width=3.0in, bb=15 0 285 230]{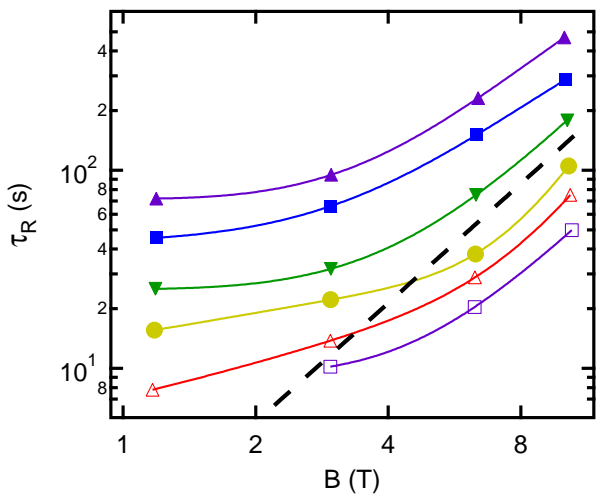}
\caption{\label{}(color online) Thermal relaxation time $\tau_R$ {\it vs.} magnetic field $B$ at several temperatures $T$. From top to bottom, $T =50$, 60, 70, 80, 90, and 100 mK.  Solid lines are guides to the eye. Clearly, the magnetic field dependence of $\tau_R$ does {\it not} obey a simple power law. The dashed line illustrates a $B^2$ dependence.}
\end{figure}
Although the origin of the long thermal time constants in our device has not yet been unambiguously determined, nuclear moments in the In-Sn contacts and the manganin wires stand out as potential culprits.  At high magnetic fields and low temperatures the nuclear spin heat capacity of In and manganin scales as $B^2/T^2$ and dwarfs that of the GaAs lattice phonons in the sample bar \cite{nuclear}.  Since these contacts and wires cool primarily through the phonon thermal conductance $K \sim T^{-2.6}$ of the sample bar, a thermal time constant $\tau \sim B^2/T^{4.6}$ results.  While this is a somewhat stronger temperature dependence than we observe ($\tau_R \sim T^{-3.7}$), the lumped ``RC'' thermal model it is based on is highly oversimplified.  

The above nuclear spin heat capacity model suggests that $\tau_R$ should vanish at $B=0$, in conflict with our observations.  However, it is well known \cite{Pobell} that manganin has a very large nuclear quadrupole heat capacity at $B = 0$ which also scales as $T^{-2}$ for $T \lesssim 0.5$ K.  This heat capacity would lead to an additional contribution to the thermal time constant of our device which again scales as $T^{-4.6}$, but now at $B=0$.  Rough estimates of the magnitude of the thermal time constant resulting from these various nuclear moments are in order-of-magnitude agreement with our observations. 

\subsection{Data Acquisition and Analysis}

As discussed in Section II-C, a quasi-DC technique is used to record the thermoelectric voltages generated by the 2DES when a temperature gradient is imposed.  At relatively high temperatures, where the thermal relaxation time of the sample bar is short, this procedure is quite straightforward.  The voltage $V_{DC}$ between two ohmic contacts is continuously recorded using a low-noise DC amplifier \cite{EM} while the heater is toggled on and off periodically. Figure 3 illustrates this with data acquired at $B = 8$ T (where $\nu = 3/2$) and $T = 180$ mK.  Under these conditions the thermal relaxation time $\tau_R \lesssim 5$ sec.  This is much smaller than the dwell time $t_D = 200$ sec. that the heater is in the on and off states in this example.  In the regime where $\tau_R << t_D$ the thermoelectric voltage $\Delta V$ is simply taken to be the difference between the averages, over several cycles, of the observed DC voltage with the heater on and off.  In effect, this technique amounts to {\it ex post facto} lock-in detection at frequencies in the mHz domain.

The thermal relaxation time $\tau_R$ grows rapidly as the temperature is reduced. For example, at $B = 8$ T and $T = 50$ mK $\tau_R \approx 300$ sec.  While one could in principle simply increase the dwell time $t_D$ of the heater in the on and off states until $t_D >> \tau_R$, the $1/f$ noise of the amplifier makes this impractical.  Instead we are forced to operate in a regime where $t_D$ is at most a few times $\tau_R$.  In this case it becomes essential to ignore those thermoelectric voltages $V_{DC}$ recorded before the sample has reached equilibrium.  We therefore exclude from the data averaging those measurements of $V_{DC}$ acquired in a specific interval following a change in the heater state.  The choice for the duration of this interval is based on our independent determinations of $\tau_R$.


\begin{references}
\bibitem{StrainGauge} We use GE/IMI 7031 varnish to attach a Vishay WK-06-031CF-350 strain gauge to our sample.
\bibitem{Q} $Q$ was always kept small compared to the cooling power of our dilution refrigerator.
\bibitem{velocity} The value of $v_{ph}$ used here, 3300 m/s, is close to the Debye velocity determined from the specific heat.
\bibitem{no_gating} After illumination the 2D density in our sample cannot be modified by gating.  This is a common occurrence in samples of this type which are intentionally over-doped.
\bibitem{nuclear} We specifically mention the In-Sn contacts and the manganin wires because they are in good thermal contact with the 2DES and have short nuclear spin-lattice relaxation times, $T_1$.  The spin 3/2 Ga and As nuclei in the sample also present a nuclear heat capacity, and may be involved as well.  Presumably, however, only those Ga and As nuclei in the quantum well containing the 2DES would have short enough $T_1$ times to be relevant.  There are far fewer such Ga and As nuclei than there are spin 9/2 In nuclei in the ohmic contacts.
\end{references}
\end{document}